 \documentclass[authoryear,preprint,review,12pt]{elsarticle}

\usepackage{amssymb}
\usepackage{graphicx}
\usepackage{caption}
\usepackage{amsmath}

\def\astrobj#1{#1}
\journal{New Astronomy}

\begin{document}

\begin{frontmatter}



\title{First Light Curve Analyses of Binary Systems \astrobj{AO~Aqr}, \astrobj{CW~Aqr} and \astrobj{ASAS~012206-4924.7}}


\author[itk]{B. Ula\c{s}\corref{dip}}
\author[abc,xyz]{C. Ulusoy}
\address[itk]{\.{I}zmir Turk College Planetarium, 8019/21 sok., No: 22, \.{I}zmir, Turkey}
\address[abc]{College of Graduate Studies, University of South Africa, PO Box 392, Unisa, 0003 Pretoria, South Africa}
\address[xyz]{Girne American University, School of Aviation, PO Box 388 , Girne, Cyprus}
\cortext[dip]{Corresponding author \\
E-mail address: bulash@gmail.com}

\begin{abstract}
Using the data from the public database of the All Sky Automated Survey ({\tt ASAS}) we performed the very first light curve analyses of the three eclipsing binary systems \astrobj{AO~Aqr}, \astrobj{CW~Aqr} and \astrobj{ASAS~012206-4924.7}. The physical parameters of the systems were determined by the {\tt PHOEBE} \citep{prs05} software. From an analysis of the ASAS data it was concluded that AO~Aqr was found to be a contact binary system while CW~Aqr and ASAS~012206-4924.7 were found to be near--contact and detached binaries, respectively.
Finally, the locations of the components, corresponding to the estimated physical parameters, in the HR diagram were also discussed.
\end{abstract}

\begin{keyword}
stars: binaries: eclipsing --- stars: fundamental parameters --- stars: individual: (\astrobj{AO~Aqr}, \astrobj{CW~Aqr},  \astrobj{ASAS~012206-4924.7})
\end{keyword}

\end{frontmatter}

\section{Introduction}

The All Sky Automated Survey ({\tt ASAS}, \cite{poj97}) project started in 1996 with the observations in the Southern hemisphere aimed at classification and investigation about 10$^{7}$ stars that show photometric variability in their light curves. The observations were carried out with the CCD cameras having 4K and 2K resolutions in the $V$ and $I$ filters at the stations located in Chile and Hawaii. The main scope of the survey was to form the ASAS Catalog of Variable Stars (ACVS) and 39000 new variables were discovered so far. 

\cite{bra80} included AO~Aqr into their catalogue of eclipsing binary stars. They listed the fundamental parameters of the components and classified the system as F type with uncertainty. \cite{giu83}  categorized AO~Aqr to be a contact binary system which mostly show EW-- and EB--type light curves in a statistical study of 1000 eclipsing binary systems. \cite{dem96} determined the accurate position of the system by using the right ascension and declination values. \cite{mal06} grouped the system as EW--type with an orbital period of 0.6984 and the spectral type of F. A variable classification catalogue based on the Northern Sky Variability Survey was published by \cite{hof09}. The same authors classified the star as W~UMa type  with an orbital period, position and magnitude values in different filters.  

The star CW~Aqr was first mentioned as a variable by \cite{hof33}. The author  provided some basic properties of 115 variable stars and noted that CW~Aqr is a binary system with a short orbital period. The system was also classified as near--contact binary with spectral type of A9 by \cite{sha94}. \cite{mal06} included the system into their catalogue of eclipsing binaries. The evolutionary status of the system was found to be near--contact  showing EB-type light curve according to their classification.

Although no other study focusing on the star ASAS~012206-4924.7 was available in the literature the fundamental parameters of the system (orbital period, a time of minimum light and magnitude in the V filter) were given by the ASAS database (Table~\ref{tablis}). The system is also included as eclipsing detached binary in the ASAS catalogue.

In this paper we present the details of the first analyses of 
the ASAS light curves in the next section. The light elements and absolute parameters are given in Sect. 3. We briefly discuss our results and the evolutionary properties of the systems in the last section.

\begin{table*}
\begin{center}
\scriptsize
\caption{Properties of the systems given by the ASAS database. Magnitudes are given in the $V$ filter. RA and DEC stand for the right ascension and the declination for the systems.}
\label{tablis}
\begin{tabular}{lcccc}
\hline
System  & RA (h m s)& DEC (d m s)& Orbital Period (days)& Magnitude (mag.) \\
\hline
AO~Aqr      &22 11 32 & -22 47 12 	&0.48934  &11.12\\
CW~Aqr      &22 19 22 & -16 53 36   &0.542913 &10.64\\
ASAS~012206-4924.7 & 01 22 06 & -49 24 42 & 2.225825 & 10.56 \\
\hline
\end{tabular}
\end{center}
\end{table*}

\section{Light Curve Analyses}
We performed the analyses of the light curves by using the {\tt PHOEBE} software \citep{prs05}. The software, which is based on the Wilson--Devinney code \citep{wil71}, applies the best fit with differential corrections to obtain the right values of the desired parameters for a given binary system. In this study, public data available in the ASAS Catalogue of Variable Stars\footnote{http://www.astrouw.edu.pl/asas/} were used to determine the physical parameters of the selected binaries.The ASAS data contain grade C and grade D points that are mentioned as "not measured" and "useless" in the header of the data files, therefore, we excluded these points from our solutions.

\subsection{\astrobj{AO~Aqr (\astrobj{ASAS~221132-2247.2})}}
To determine the physical parameters of AO~Aqr we derived the effective temperature of the primary component from the value of $B-V$=$0.64$ mag given in the {\tt SIMBAD} database \citep{wen00}. Therefore, we assumed the temperature $T_1$=$5750$~K by using the correlation given by \cite{flo96}. The result is in agreement with the other calculations given by \cite{bra80} and \cite{mal06}. The spectral type of the system was determined as F by the same authors. For the light curve solution of AO~Aqr the time of minimum light $T_0$, the orbital period of the binary $P$, mass ratio $q$, inclination $i$, temperature of secondary component $T_2$, surface potential $\Omega_1 = \Omega_2$ and luminosity of the primary component $L_1$ were set as free parameters. The albedo values A$_1$,A$_2$ were adopted from \cite{ruc69} while the gravitational darkening coefficients g$_1$, g$_2$ were derived from the values given by \cite{ham93}. The filling factor of the binary was found to be $f$=74\% using the formula $f=\frac{\Omega_{i}-\Omega}{\Omega_{i}-\Omega_{o}}$, where $\Omega_{i}$ is the inner and $\Omega_{o}$ is the outer Lagrangian potentials.  Table~\ref{tablc} lists the values of parameters for the best fits during the analysis. The light curve with the theoretical fit is plotted in Fig.~\ref{lcgeo}.

\begin{table}
\scriptsize
\caption{Result of the light curve analysis.}
\label{tablc}
\begin{tabular}{lccc}
\hline
Parameter  & AO~Aqr & CW~Aqr & ASAS~012206-4924.7  \\
\hline
$i$ ${({^\circ})}$              & 79.4(1)  &73.4(4) &86.4(2)\\
$q$                             & 0.288(6) &0.252(8) &0.59(2)\\
$\Omega _{1} $                  & 2.31(1) & 2.43(2) &7.42(2)  \\
$\Omega _{2} $				&$\Omega _{1}$ &$\Omega _{cr}=2.35$ &5.34(1) \\
$T_1$~(K)                   & 5750 &7700 & 6000  \\
$T_2$~(K)                               & 5708(33) & 5018(81) &5783(34)\\
$\overline{r_{1}}$              & 0.530(6) & 0.482(8) &0.147(7)\\
$\overline{r_{2}}$              & 0.325(20)& 0.266(4) &0.144(9)\\
$\frac{L_1}{L_1 +L_2 +l_3}$ &  0.753(7)  & 0.957(9)    & 0.56(3) \\
$T_0$~(HJD-2450000)       & 4680.8776(83)   &  2144.6767(89) &2947.6559(156) \\
$P$~(days)         & 0.489352(1)  & 0.542908(2)  & 2.225825(12)\\ 
\hline
\end{tabular}
\end{table}

\subsection{\astrobj{CW~Aqr (\astrobj{ASAS~221922-1653.6})}}

For CW~Aqr, we followed the same steps explained as in the light curve solution of AO~Aqr. However, the temperature of the primary component was calculated by considering the spectral type (A9V- assuming the star is on the ZAMS) given by \cite{sha94}. The temperature value of the hotter component was estimated to be 7700~K that was fitted to the spectral type using the correlation given by \cite{cox00}. In addition, since the system was hypothesized to be semi--detached binary the analyses were performed in the suitable mode of the {\tt PHOEBE} software. Therefore, the surface potential value of the secondary component $\Omega_2$ was set to the critical potential value of the first Lagrangian point as distinct from the solution of AO~Aqr. The light curve fitting is plotted in Fig~\ref{lcgeo}, and the resulting parameters are given in Table~\ref{tablc}. 

\subsection{\astrobj{ASAS~012206-4924.7}}

\begin{figure*}
\begin{center}
\includegraphics[scale=0.8]{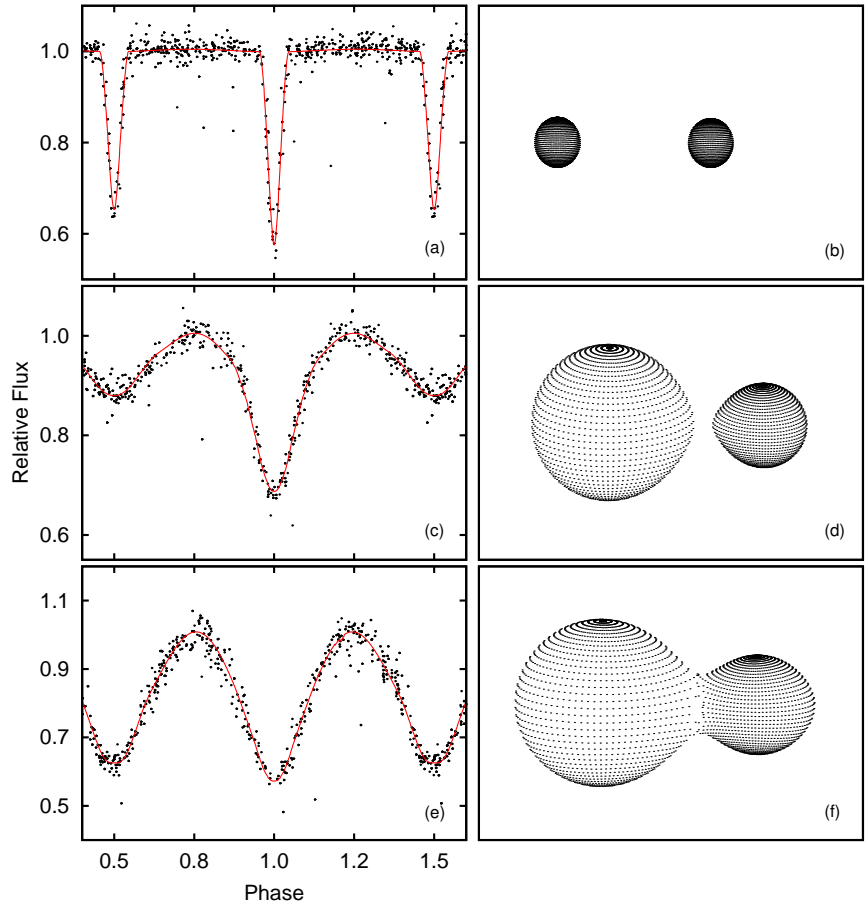}
\caption{The theoretical light curves are compared to observations for ASAS~012206-4924.7 (a), CW~Aqr (c) and AO~Aqr (e). The geometric configurations of ASAS~012206-4924.7 (b), CW~Aqr (d) and AO~Aqr (f) at 0.25 phase are drawn according to our results}
\label{lcgeo}
\end{center}
\end{figure*}

The light curve analysis applied to the data by following the similar steps was explained in the previous subsections. Since the system is suspected to be a detached binary we set the potential values $\Omega_1$ and $ \Omega_2$ as free parameters in the detached binary mode of the program. The temperature of the primary component was adopted from \cite{flo96} for the correct $B-V$ value by using the magnitudes given in the {\tt SIMBAD} database. The results can be seen in Table~\ref{tablc}. The light curves yielded by the solution are presented in Fig~\ref{lcgeo}.

\section{Discussion and Conclusion}

We present the light curve solutions of the three binary systems selected from the ASAS database. Our analyses are the first analyses of the systems in the literature. The absolute parameters of the systems were determined by using the results of the light curve solution and listed in Table~\ref{tababs}. We used the calibration between spectral type and mass values given by \cite{cox00} in order to calculate the masses of the primary components.

\begin{figure*}
\begin{center}
\includegraphics[scale=0.7]{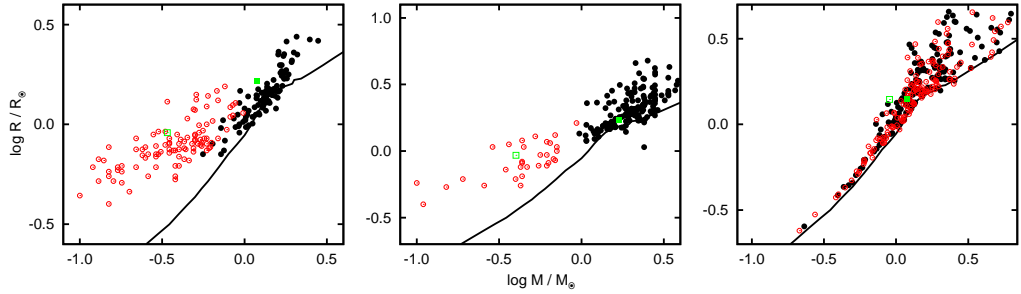}
\caption{The position of the components of the systems in different mass-radius planes. The left panel shows the agreement between the components of AO~Aqr to other 100 contact binaries while 31 near--contact systems are located with the components of CW~Aqr in the middle panel. The right panel is a plot of the components of ASAS~012206-4924.7 with 168 detached systems. Filled and open circles indicate the primary and secondary components of the systems taken from literature. The primary and secondary components of the targets, on the other hand, are represented by filled and open squares. ZAMS data are taken from \cite{gir00}.}
\label{figmr}
\end{center}
\end{figure*}

\begin{figure}
\includegraphics{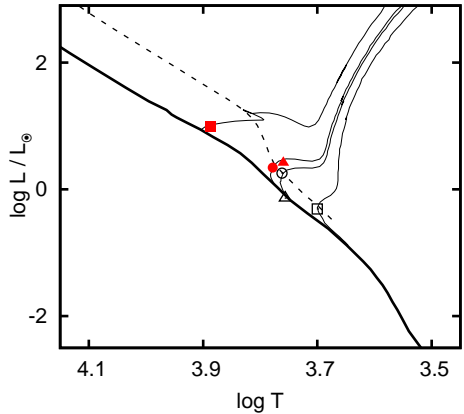}
\caption{Location of the components of the systems on the Hertzsprung--Russell diagram. Filled symbols refer to primary components while the empty symbols denote the secondaries. The triangles are used to symbolize the contact binary AO~Aqr. The squares indicate the components of semi--detached binary CW~Aqr. The circles show the detached system ASAS~012206-4924.7. The evolutionary tracks represent the evolution for the stars having initial masses of  1.7, 1.1, 1.0 and 0.7 M$_{\odot}$ from left to right. Finally, the thick solid line and the dashed line stand for ZAMS and TAMS, respectively. The evolutionary tracks for Z=0.019 and ZAMS data are taken from \cite{gir00}.}
\label{fighr}
\end{figure}

AO~Aqr is a contact binary system. Contact binary systems are divided into two subclasses (A-- and W-- type W~UMa) by \cite{bin70} according to morphology of their light curves. The system is found to be an A--type W~UMa based on this classification since the more massive component is occulted during the primary eclipse. An important evolutionary indication of these systems were made by \cite{luc68}. Recently, \cite{ste12} developed a model for cool contact binaries. They remarked that the systems evolve from detached binaries that have orbital period of about 2 days. As the massive star expands it fills the Roche lobe and starts to transfer mass to the less massive component. When the mass ratio reverse the rapid mass transfer ends and the contact system is formed by the contribution of angular momentum loss. It should be noted that angular momentum plays an important role in the case of evolution of contact phase. Therefore, the evolution ends in a rapidly rotating single star formed by the process of merging of the components. In our study we compare our parameters obtained by the analysis of AO~Aqr with the parameters of the contact binaries on mass--radius plane (Fig.~\ref{figmr}). The results are in a good agreement with the other 100 contact systems given by \cite{yil13}. The location of the components, corresponding to the determined physical parameters, in the HR diagram is shown in Fig.~\ref{fighr}. The primary component is very close to the TAMS while the secondary one is located slightly below the ZAMS.

The filling factor of the primary component ($f=r/r_L$) of CW~Aqr was found to be 96\% using the volume radius of the Roche lobe given by \cite{egg83}:
\begin{equation}
r_L = \frac{ 0.49q^{\frac{2}{3}} } {0.6q^{\frac{2}{3}} + \ln(1+q^{\frac{1}{3}})}
\end{equation}
Therefore, the component is very close to the Roche lobe which proves that the binary is a near--contact system as listed by \cite{sha94}. Our results show that the system can be considered as a member of the FO~Virginis class according to the near--contact classification of \cite{sha94} since the secondary component is at its Roche lobe and the light curve does not show any noticeable difference between the two maxima. \cite{sha94} proposed that these type of binaries would be the progenitors of A--type W~UMa systems. CW~Aqr is shown on the mass-radius plane with other 31 near--contact systems whose parameters were adopted from \cite{hil88} in Fig.~\ref{figmr}. The  components of the system were placed in the same position as expected from a near--contact binary. The location of the primary and secondary components in the HR diagram indicate that both components are close to the ZAMS. The evolutionary tracks of the stars having initial mass values of 1.7 M$_{\odot}$ and 0.7 M$_{\odot}$ overlap with the locations of primary and secondary components (slightly above the ZAMS),respectively.

As aforementioned above, the cooler contact binaries evolve from the detached systems with period of about 2 days in the binary model of \cite{ste12}. Accordingly, the star ASAS~012206-4924.7, with the period of about 2.2 days, can be considered as a candidate in the similar scenario. As can be seen in Fig.~\ref{figmr}, The position of the components is in a good agreement with the other 160 detached binaries cataloged by \cite{sou14}. In the HR diagram (Fig.~\ref{fighr}), the primary and secondary components are situated on the same evolutionary tracks of the stars with initial masses of 1.1 M$_{\odot}$ and 1.0 M$_{\odot}$ .

\begin{table}
\scriptsize
\caption{Absolute parameters of the system derived from the light curve solution results. The effective temperature of the sun is set to 5780~K.}
\label{tababs}
\begin{tabular}{lcccccc}
\hline 
Parameter                &\multicolumn{2}{c}{AO~Aqr}  &\multicolumn{2}{c}{CW~Aqr} &\multicolumn{2}{l}{ASAS~012206-4924.7} \\
& \multicolumn{1}{c}{P}&\multicolumn{1}{c}{S.} & \multicolumn{1}{c}{P}&\multicolumn{1}{c}{S} & \multicolumn{1}{c}{P}&\multicolumn{1}{c}{S} \\
\hline
M (M$_{\odot}$)  & 1.19         & 0.34(1) & 1.68     & 0.4(2)  &1.19 &0.9(1) \\
R (R$_{\odot}$)  & 1.64(2)      & 0.9(1)  & 1.7(1)   & 0.93(6) &1.4  & 1.4(2)\\
L (L$_{\odot}$)  & 2.62(6)     & 0.7(2)   & 9.9(1.3) & 0.49(7) &2.2(5) & 1.8(5)\\
T (K)			 & 5750		    & 5708(33) & 7700    & 5018(81) &6000 & 5783(33)\\
$a$ (R$_{\odot}$)  &\multicolumn{2}{c}{3.1(1)}  &\multicolumn{2}{c}{3.7(1)} &\multicolumn{2}{c}{9.5(4)} \\
\hline
\end{tabular}
\end{table}

\section*{Acknowledgement}

CU acknowledges financial support from the University of South Africa
(UNISA) and the South African National Research Foundation (NRF), Grant
No:87635. CU also wishes to thank Prof. Greg Cuthbertson, Prof.
Michele Havenga and Mr. Prince Tatz Nyoka for their valuable contribution to her research. This research has made use of the SIMBAD database, operated at CDS, Strasbourg, France.


\begin{thebibliography}{}

\bibitem[\protect\citeauthoryear{Brancewicz \& Dworak}{1980}]{bra80} Brancewicz H.~K., Dworak T.~Z., 1980, AcA, 30, 501
\bibitem[\protect\citeauthoryear{Binnendijk}{1970}]{bin70} Binnendijk L., 1970, VA, 12, 217
\bibitem[\protect\citeauthoryear{Cox}{2000}]{cox00} Cox A.~N., 2000, asqu.book
\bibitem[\protect\citeauthoryear{Demartino et al.}{1996}]{dem96} Demartino R., Kocyla D., Predom C., Wetherbee E., 1996, IBVS, 4321, 1 
\bibitem[\protect\citeauthoryear{Eggleton}{1983}]{egg83} Eggleton P.~P., 1983, ApJ, 268, 368 
\bibitem[\protect\citeauthoryear{Flower}{1996}]{flo96} Flower P.~J., 1996, ApJ, 469, 355 
\bibitem[\protect\citeauthoryear{Giuricin, Mardirossian, \& Mezzetti}{1983}]{giu83} Giuricin G., Mardirossian F., Mezzetti M., 1983, A\&AS, 54, 211
\bibitem[\protect\citeauthoryear{Girardi et al.}{2000}]{gir00} Girardi L., Bressan A., Bertelli G., Chiosi C., 2000, A\&AS, 141, 371
\bibitem[\protect\citeauthoryear{Hoffman, Harrison, \& McNamara}{2009}]{hof09} Hoffman D.~I., Harrison T.~E., McNamara B.~J., 2009, AJ, 138, 466 
\bibitem[\protect\citeauthoryear{Hilditch, King, \& McFarlane}{1988}]{hil88} Hilditch R.~W., King D.~J., McFarlane T.~M., 1988, MNRAS, 231, 341 
\bibitem[\protect\citeauthoryear{Hoffmeister}{1933}]{hof33} Hoffmeister C., 1933, AN, 247, 281
\bibitem[\protect\citeauthoryear{Lucy}{1968}]{luc68} Lucy L.~B., 1968, ApJ, 151, 1123 
\bibitem[\protect\citeauthoryear{Malkov et al.}{2006}]{mal06} Malkov O.~Y., Oblak E., Snegireva E.~A., Torra J., 2006, A\&A, 446, 785 
\bibitem[\protect\citeauthoryear{Pojmanski}{1997}]{poj97} Pojmanski G., 1997, AcA, 47, 467 
\bibitem[\protect\citeauthoryear{Pr{\v s}a \& Zwitter}{2005}]{prs05} Pr{\v s}a A., Zwitter T., 2005, ApJ, 628, 426
\bibitem[\protect\citeauthoryear{Ruci{\'n}ski}{1969}]{ruc69} Ruci{\'n}ski S.~M., 1969, AcA, 19, 245
\bibitem[\protect\citeauthoryear{Shaw}{1994}]{sha94} Shaw J.~S., 1994, MmSAI, 65, 95
\bibitem[\protect\citeauthoryear{Southworth}{2014}]{sou14} Southworth J., 2014, arXiv, arXiv:1411.1219 
\bibitem[\protect\citeauthoryear{St{\c e}pie{\'n} \& Gazeas}{2012}]{ste12} St{\c e}pie{\'n} K., Gazeas K., 2012, AcA, 62, 153
\bibitem[\protect\citeauthoryear{van Hamme}{1993}]{ham93} van Hamme W., 1993, AJ, 106, 2096
\bibitem[\protect\citeauthoryear{Wenger et al.}{2000}]{wen00} Wenger M., et al., 2000, A\&AS, 143, 9
\bibitem[\protect\citeauthoryear{Wilson \& Devinney}{1971}]{wil71} Wilson R.~E., Devinney E.~J., 1971, ApJ, 166, 605
\bibitem[\protect\citeauthoryear{Yildiz \& Do{\u g}an}{2013}]{yil13} Yildiz M., Do{\u g}an T., 2013, MNRAS, 430, 2029

\end{thebibliography}
\end{document}